\documentclass[]{jsedi_pub}

\title{The core-mantle mode of gravitational oscillation}
\shorttitle{core-mantle gravitational oscillation} 

\author[1]{Mathieu Dumberry
	\orcid{0000-0001-5677-1582}
	\thanks{Corresponding author: 
    \href{mailto:dumberry@ualberta.ca}{dumberry@ualberta.ca}}
}

\affil[1]{Department of Physics, University of Alberta, Edmonton AB, Canada}

\begin{document}

\publicationonly{
\dois{10.46298/jsedi.17576}
\handedname{Alexandre Fournier}
\receiveddate{February 24, 2026}
\reviseddate{May 25, 2026; July 31, 2026}
\accepteddate{August 26, 2026}
\publisheddate{September 4, 2026}
\theyear{2026}
\thevolume{2}
\thepaper{7}  
}

\makesedititle{
  \begin{summary}{Abstract}
We present the characteristics of a mode of axial oscillation between Earth's mantle, fluid core and solid inner core that has not been previously reported.  The mode involves a quasi-rigid rotation of the fluid core outside the tangent cylinder (TC) exchanging its angular momentum with the mantle via a three step process. First, by a magnetic torque with the fluid inside the TC; second by a magnetic torque between the latter and the inner core; and finally by a gravitational torque between the inner core and mantle.  Although the gravitational torque is purely between the mantle and inner core, the mode involves an oscillation of the whole of the core, and we refer to it as the core-mantle gravitational (CMG) mode. This form of gravitational oscillation occurs when the magnetic field within the core is sufficiently strong that the propagation time of Alfv\'en waves is shorter than the mode period, which is the case for Earth.  We show how the period, quality factor $Q$ and structure of the CMG mode depends on the strength of the gravitational torque and viscous relaxation time $\tau_i$ of the inner core.  For Earth, the CMG mode period should be in the range of 40 to 100 years, but viscous relaxation of the inner core likely implies a small $Q$, below 1 if $\tau_i<10$ years. Our results suggest that the CMG mode may act to amplify resonantly, though only modestly, multi-decadal changes in the length of day driven by zonal accelerations in the fluid core.
    \vspace{1.3cm} 
  \end{summary}
  }
 \begin{summary}{Non-technical summary}
The Earth's mantle, fluid core and inner core are coupled by forces that allow for natural modes of oscillations between them.  Here, we describe a mode that we term the core-mantle gravitational mode, or CMG mode for short.  The mode is sustained by the gravitational torque between the inner core and mantle and by the internal magnetic field that couples the inner core to the fluid core.  We compute the period, the decay rate, and the structure of the motion in the fluid core for different sets of parameters.  We show that, based on current estimates of the strength of the gravitational coupling, the period of the CMG mode should be in the range of 40 to 100 years.  However, we also show that viscous relaxation of the inner core and electromagnetic friction at the core-mantle boundary attenuate the CMG mode rapidly.  This suggests that it is unlikely that the CMG mode can account for a specific frequency in the spectrum of the observed decadal changes in the length of day. However, it is possible that the presence of the CMG mode may help to amplify resonantly, though only modestly, multi-decadal changes in the length of day driven by zonal accelerations in the fluid core.
 \end{summary}

\section{Introduction}

To first order, mantle convection on Earth is characterized by a circum-Pacific ring of cold, downwelling subducted plates separating two antipodal regions of upwelling, one under the Pacific ocean and the other under Africa \citep[e.g.][]{defraigne96,simmons07}. This produces a mass anomaly primarily at spherical harmonic degree 2 order 2 reflected in the observed geoid at the surface \citep{hager85}.  This mantle mass anomaly imposes a gravitational potential within the core, deforming the inner core boundary (ICB) topography into an equatorially elliptical shape \citep{buffett96a}.  Because of the density contrast between the inner core and the fluid core, this ICB topography creates an additional mass anomaly. 

If the inner core is axially rotated, an axial gravitational torque acts to restore the longitudinal alignment between the ICB and mantle mass anomalies.  This torque provides the restoring force for a natural mode of axial oscillation \citep{buffett96a} which has been referred to as the mantle-inner core gravitational (MICG) mode \citep{mound06}. Electromagnetic (EM) coupling between the fluid and inner core at the ICB should be strong \citep[e.g.][]{gubbins81}, and as a result the region of the fluid core inside the tangent cylinder (TC, i.e. the axial cylinder tangent to the equator of the inner core) is expected to be entrained into co-rotation with the inner core \citep{mound06}.  The period of the MICG mode depends on the amplitude of the degree 2 order 2 mantle mass anomalies  \citep{buffett96a,davies14,chao17,shih21} and is estimated to be in the range of 7-18 years \citep{davies14}.

How the region of the fluid core outside the TC (ROTC) affects the MICG depends on the strength of the magnetic field within the core.  At decadal periods, axisymmetric azimuthal (zonal) motion in the ROTC takes the form of rigidly rotating nested cylindrical surfaces \citep{jault08}.  Adjacent cylinders are coupled by the magnetic field that threads their surfaces.  This coupling allows for the propagation of Alfv\'en waves in the cylindrically radial direction and for normal modes of axial oscillations between cylinders referred to as torsional oscillations \citep[TO,][]{braginsky70}. For a weak magnetic field of the order of 0.2-0.3 mT -- once believed to be a representative strength within the core -- the period of the fundamental mode of TO is in the range of 60-80 years \citep{zatman97}, much longer than the MICG mode period.  The faster oscillations of the TC involved in the MICG mode launch Alfv\'en waves in the ROTC that excite high harmonics of TO modes with short wavelengths, and large cancellations occur in the axial angular momentum carried by the ROTC \citep{mound03}. Consequently, when the magnetic field is weak, the ROTC plays a limited role in the angular momentum dynamics of the MICG mode and the MICG period is virtually identical to that computed under the assumption of fully decoupled ROTC \citep{mound03,mound06}. However, for a magnetic field strength within the core of a few mT -- as we now believe to be the case \citep{gillet10} --  the period of the fundamental mode of TO is approximately 6 years.  In this case, the similar period of the MICG implies that the oscillating TC launches Alfv\'en waves that excite low harmonics of TO modes with long wavelengths and the ROTC carries a significant net axial angular momentum \citep[][henceforth referred to as Paper 1]{dumberry25}.  As a result, the oscillating motion of the TC involved in the MICG mode cannot be decoupled from these TO modes.  Indeed, Paper 1 showed that when Alfv\'en waves can readily traverse the width of the core before they get attenuated, the MICG mode is no longer an individual mode of the coupled core-mantle system; instead, it gets absorbed into the spectrum of TO modes.

\begin{figure*}[ht!]
  \centering
  \includegraphics[width=0.9\textwidth]{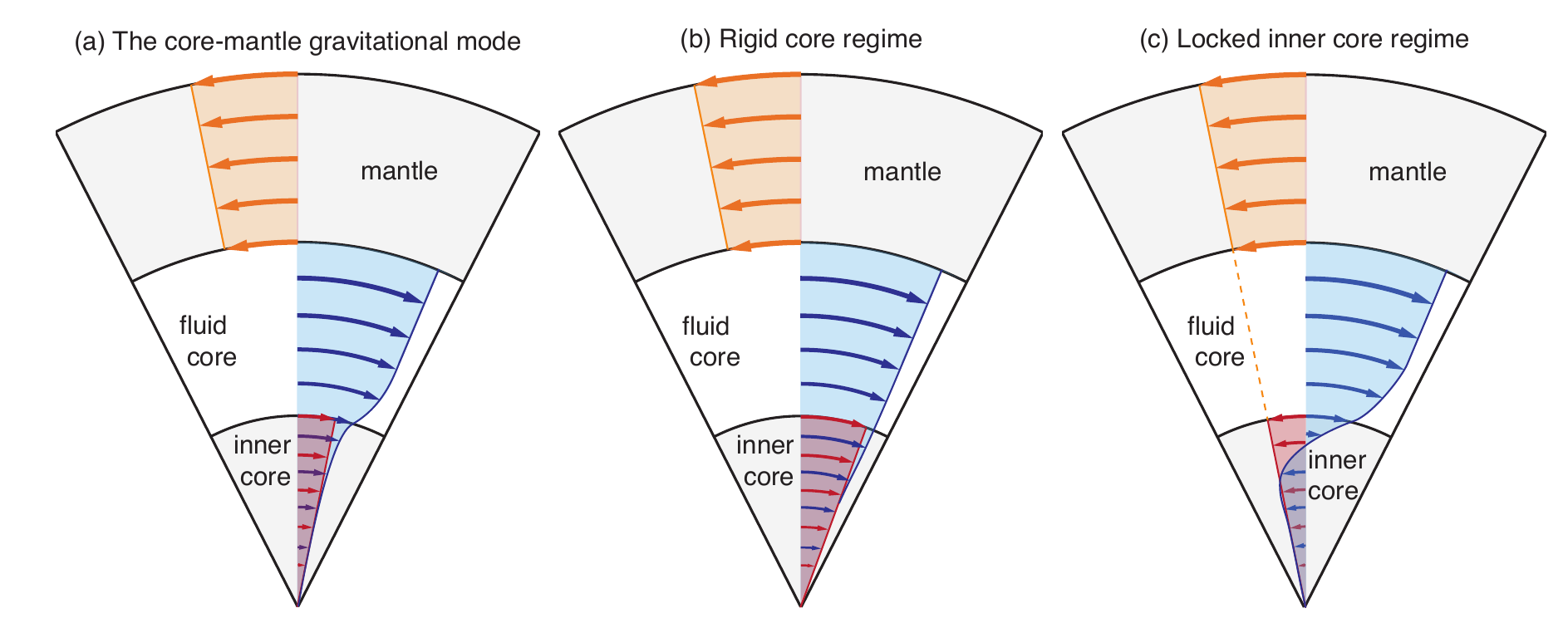} 
  \caption{The angular velocity structure of the mantle ($\Omega_m$ orange arrows), fluid core ($\Omega_f$ blue arrows) and inner core ($\Omega_i$ red arrows) in (a) a generic case of the CMG mode, (b) the rigid core regime and (c) the locked inner core regime.  The part of the fluid core overlapping the inner core represents the flow inside the TC.  For ease of visualization, the relative angular velocity amplitudes and the radii of each region are not drawn to scale.}
  \label{fig:cmg_cartoon}
\end{figure*}

However, as we shall show, a different mode of gravitational oscillation between the mantle and the core is possible in a strong field regime. When the angular momentum of the core changes slowly compared to the propagation time of Alfv\'en waves, the latter act to eliminate differential angular zonal motion in the ROTC.  That is, the magnetic field effectively locks the whole of the ROTC into a quasi-rigidly rotating body. It is then possible to have a global mode of axial oscillation which involves an EM coupling at the TC between this rigidly rotating ROTC and the fluid region inside the TC (RITC), an EM coupling between the RITC and the inner core at the ICB, and a gravitational coupling between the inner core and mantle.  Hence, a gravitational mode of oscillation between the mantle and inner core remains possible, though one that involves the whole of the core instead of only the TC (see Figure \ref{fig:cmg_cartoon}a).  Since the moments of inertia of the inner core and the RITC are much smaller than that of the ROTC, the angular momentum is conserved primarily through a balance between the oppositely oscillating ROTC and mantle.  To distinguish this mode from the MICG, we refer to it as the core-mantle gravitational (CMG) mode.  As we show below, the period of the CMG mode is longer than that of the MICG mode, in the multi-decadal range.

The objective of the present study is to present the characteristics of the CMG mode.  As we describe below, whether the inner core oscillates in sync with the ROTC or instead with the mantle depends on the relative strengths of the gravitational coupling between the inner core and mantle versus the EM coupling between the RITC and ROTC.  When the EM coupling strength at the TC is much larger, the inner core follows the ROTC and the entire core rotates as a quasi-rigid body (Figure \ref{fig:cmg_cartoon}b);  we refer to this limit as the rigid core regime. When instead it is the gravitational coupling that is stronger, the motion of the inner core remains quasi-locked to the mantle (Figure \ref{fig:cmg_cartoon}c); we refer to this limit as the locked inner core regime.  

Our study is organized as follows.  In section 2, we show how the CMG mode emerges from the angular momentum equations of the core-mantle system.  We present analytical expressions for its period and quality factor $Q$ in both the rigid core and locked inner core regimes.  In section 3, we compute numerically the CMG mode and show how its period and $Q$ depend on the strength of the gravitational coupling factor between the inner core and mantle, the viscosity of the inner core, and EM damping at the CMB.  We conclude our study in section 4 with a geophysical discussion.

\section{Theory}

\subsection{The coupled core-mantle system of axial oscillations}

Paper 1 presents a model that captures the coupled oscillations between the mantle, inner core (radius $r_i$) and fluid core (outer radius $r_f$) in the absence of an external torque.  This coupled core-mantle system broadly follows that developed in several earlier studies \citep{buffett96b,buffett98,mound03,mound05b,dumberry08c,dumberry10b}. It includes TO modes in the fluid core, EM coupling at the ICB and CMB, and gravitational coupling between the mantle and inner core. The full development of the model will not be repeated here, but in order to extract the basic attributes of the CMG mode, it is convenient to reproduce the equations that govern its angular momentum dynamics.

We use $C_m$, $C_i$ to denote the axial moments of inertia of the mantle and inner core, and $\Omega_m$, $\Omega_i$ to denote the fluctuations of their angular velocities with respect to their mean rotation rates.  Temporal changes in the axial angular momentum of the fluid core are carried by zonal flows in the form rigid cylindrical surfaces aligned with the rotation axis \citep[e.g.][]{jault08}. We use cylindrical coordinates ($s$, $\phi$, $z$), so that a cylinder with a cylindrical radius $s$ intersects the CMB at axial position $z_f=\sqrt{r_f^2 - s^2}$ in the Northern hemisphere.  Cylinders inside the TC ($s<r_i$) intersect the ICB at $z_i=\sqrt{r_i^2 - s^2}$.  We use $\Omega_f=\Omega_f(s)$ to denote the fluctuation of the angular velocity of a cylinder located at cylindrical radius $s$ with axial moment of inertia density
\begin{equation}
c_f = 4 \pi \rho s^3 (z_f - z_i) \, ,
\end{equation}
where $\rho$ is the density and where $z_i=0$ for $s>r_i$.

The axial angular momentum equations for the mantle and inner core are given, respectively, by
\begin{subequations}
\begin{align}
 C_m \frac{d\Omega_m}{dt}   & =  \overline{\Gamma} \alpha +  \int_0^{r_f}  {\cal F}_m  \left[ \Omega_f - \Omega_m \right] \, ds  \, , \label{eq:tqm} \\
 C_i  \frac{d\Omega_i}{dt} & = - \overline{\Gamma} \alpha +   \int_0^{r_i} {\cal F}_i  \left[ \Omega_f - \Omega_i \right] \,  ds \, ,\label{eq:tqi} 
\end{align}
where $\overline{\Gamma}$ captures the strength of the gravitational coupling between the mantle and inner core and $\alpha$ is the longitudinal angle of misalignment between the degree 2 order 2 mantle density field and ICB topography.  The evolution of $\alpha$ is given by 
\begin{equation}
\frac{d\alpha}{dt} = \Omega_i - \Omega_m - \frac{\alpha}{\tau_i} \, ,
\label{eq:dtalpha}
\end{equation}
\end{subequations}
where $\tau_i$ is the viscous relaxation time of the inner core assuming a Maxwell rheology.  The parameters ${\cal F}_m$ and ${\cal F}_i$ capture the strength of the torques from all surface forces acting on the mantle at the CMB and on the inner core at the ICB, respectively.  We assume that ${\cal F}_m$ and ${\cal F}_i$ result from electromagnetic (EM) coupling as specified by eqs. (5a) and (5b) of Paper 1. 

The temporal fluctuations in $\Omega_f$ of individual cylinders in the fluid core are governed by 
\begin{subequations}
\begin{align} c_f \frac{\partial \Omega_f}{\partial t}  & = \frac{1}{\rho \mu} \frac{\partial }{\partial s} \left(c_f \frac{\{B_s b_\phi\}}{s} \right)  - {\cal F}_m \left[ \Omega_f - \Omega_m \right] \nonumber\\
&\quad - {\cal F}_i \left[ \Omega_f - \Omega_i \right]  + \nu \frac{\partial}{\partial s} \left(c_f \frac{\partial \Omega_f}{\partial s} \right) \, , \label{eq:omf} 
\end{align}
where $\mu$ is the magnetic permeability, $\nu$ is the kinematic viscosity, $B_s$ is the $s$-component of the background magnetic field, and $\{ \cdot\}$ denotes an average over a cylindrical surface. The variable $b_\phi$ is the azimuthal magnetic field perturbation induced by the differential motion of the cylinders, and obeys
\begin{equation}
\frac{\partial b_\phi}{\partial t}  = s B_s \frac{\partial \Omega_f}{\partial s} + \eta \left( \nabla^2  {\bf b}  \right)_\phi \, , \label{eq:bphi} 
\end{equation}
\end{subequations}
where $\eta=1/\mu \sigma$ is the magnetic diffusivity.

Eq. \eqref{eq:bphi} can be multiplied by $B_s$ and averaged over a cylindrical surface, thus forming an equation for the evolution of $\{ B_s b_\phi\}$.  The combination of eq. \eqref{eq:omf} and this modified induction equation form a reduced one-dimensional (1D) system that captures the leading order angular momentum dynamics of magnetically coupled cylinders in the core, including the propagation of Alfv\'en waves    \citep[e.g.][]{jault03,jault05,canet09,cox14,maffei16}. The induction term involves the quantity $\{B_s^2\}$, or equivalently, the rms strength of the $B_s$-field averaged over a cylindrical surface, $\{B_s\}$ = $\sqrt{\{B_s^2\}}$.  For this reduced 1D system, $\{B_s\}$ is the only part of the background magnetic field that influences the dynamics of $\Omega_f$.  It suffices then to prescribe a profile of $\{B_s\}$ as a function of $s$, with the understanding that this profile results from a fully three dimensional and complex magnetic field.  

Since $\Omega_f$ depends only on $\{ B_s\}$, an alternate form of this 1D system can be written by replacing $B_s$ with $\{B_s\}$ in eqs. \eqref{eq:omf} and \eqref{eq:bphi}, and taking the average of the latter over a cylindrical surface.  This alternate 1D system is  

\begin{subequations}
\begin{align}
c_f \frac{\partial \Omega_f}{\partial t} & = \frac{1}{\rho \mu} \frac{\partial }{\partial s} \left(c_f \frac{\{B_s\} b_\phi}{s} \right)  - {\cal F}_m \left[ \Omega_f - \Omega_m \right] \nonumber\\
& \quad- {\cal F}_i \left[ \Omega_f - \Omega_i \right] + \nu \frac{\partial}{\partial s} \left(c_f \frac{\partial \Omega_f}{\partial s} \right) \, , \label{eq:omfapp} 
\end{align}
\begin{equation}
\frac{\partial b_\phi}{\partial t}  = s \{B_s\} \frac{\partial \Omega_f}{\partial s} + \eta \left(  \frac{\partial^2 b_\phi}{\partial s^2} + \frac{1}{s} \frac{\partial b_\phi}{\partial s} - \frac{b_\phi}{s^2} \right) \, , \label{eq:bphiapp} 
\end{equation}
\end{subequations}
where $b_\phi$ now represents an average over a cylindrical surface.  This formalism has been used in a few studies \citep[e.g.][]{more18,dumberry20}, including eqs. (10) and (11) of Paper 1, and we also use it here.  Other than involving a different magnetic field variable ($b_\phi$ instead of $\{ B_s b_\phi\}$), in the absence of magnetic diffusion, the dynamics of $\Omega_f$ is identical in either form of the 1D system (see Appendix A).  The alternate 1D system of eqs. \eqref{eq:omfapp} and \eqref{eq:bphiapp} offers no specific advantage over the original one, other than that the magnetic diffusion term is perhaps easier to construct and interpret.  

One limitation of using a reduced 1D system (either the original or alternate version) is that it underestimates the effects of magnetic diffusion in the fluid core (see Appendix A).  As we will show, the CMG mode features a core flow with weak gradients in $s$ (at least in the rigid core regime) which should minimize magnetic diffusion. In this sense, the salient features of the CMG mode should be properly captured with this simple 1D model. 

Eqs. \eqref{eq:omfapp} and \eqref{eq:bphiapp} require boundary conditions at  $s=0$ and $s=r_f$.  At $s=0$, we impose the regularity conditions $d\Omega_f/ds=0$ and $b_\phi=0$.  At $s=r_f$, we set $d\Omega_f/ds=0$, which implies no viscous torque at the CMB, and $b_\phi$ is set by the magnetic field perturbation resulting from EM coupling (eq. 16 of Paper 1).  

\subsection{The core-mantle gravitational mode}

Let us now show how the CMG mode emerges from the above system.  The CMG mode can be broadly described as an axial oscillation between the mantle, inner core, and the quasi-rigid RITC and ROTC. Let us define the moments of inertia of the RITC and ROTC, respectively $C_{fi}$ and $C_{fo}$, as   
\begin{subequations}
\begin{align}
C_{fi} & =  \int_0^{r_i} c_f \, ds \, ,\label{eq:Cfi} \\
C_{fo} & =  \int_{r_i}^{r_f} c_f \, ds \, .\label{eq:Cfio} 
\end{align}
\end{subequations}
Let us further introduce the mean angular velocities inside and outside the TC, respectively $\Omega_{fi}$ and $\Omega_{fo}$, defined such that
\begin{subequations}
\begin{align}
 \int_0^{r_i} c_f \Omega_f(s) \, ds \, & = C_{fi} \, {\Omega}_{fi} \, , \label{eq:omegafi} \\
 \int_{r_i}^{r_f} c_f \Omega_f(s) \, ds \, &= C_{fo} \, {\Omega}_{fo} \, .  \label{eq:omegafo} 
\end{align}
\end{subequations}
By this construction, $\Omega_{fi}$ and $\Omega_{fo}$ carry all the angular momentum of the core.  For periodic oscillations between $\Omega_i$, $\Omega_m$, $\Omega_{fi}$ and $\Omega_{fo}$ about a mean rotation rate, conservation of angular momentum in the absence of an external torque is expressed as
\begin{equation}
C_i \Omega_i + C_{fi} \Omega_{fi} + C_{fo} \Omega_{fo} + C_m \Omega_m  = 0 \, . \label{eq:angmom}
\end{equation}
A necessary condition for the CMG mode is a strong internal magnetic field, in which case EM coupling at both the ICB and the TC should be strong. This implies that $\Omega_i$ and $\Omega_{fi}$ should not be much larger than $\Omega_{fo}$.  However, since $C_{i}, C_{fi} \ll C_{fo}$, to first order, the angular momentum balance in the CMG mode is primarily between the mantle and the ROTC,
\begin{equation}
C_{fo} \Omega_{fo} \approx - C_m \Omega_m  \, . \label{eq:angmom_approx}
\end{equation}

Evolution equations for $\Omega_{fi}$ and $\Omega_{fo}$ can be formed by integrating eq. \eqref{eq:omf} from $s=0$ to $r_i$ and from $s=r_i$ to $r_f$, respectively.  Taking the time derivative of each of the resulting equations, inserting eq. \eqref{eq:bphi}, and neglecting both viscous and magnetic diffusion within the core yields
\begin{subequations}
\begin{align}
C_{fi} \frac{\partial^2 \Omega_{fi}}{\partial t^2} &=   \frac{\{B^2_s\} }{\rho \mu} \left(c_f \frac{\partial \Omega_f}{\partial s} \right) \Bigg|_{s=r_i} - \int_o^{r_i} {\cal F}_i \frac{\partial}{\partial t} \left[ \Omega_f - \Omega_i \right] ds \nonumber\\
&\quad  - \int_o^{r_i} {\cal F}_m \frac{\partial}{\partial t} \left[ \Omega_f - \Omega_m \right] ds \, , \label{eq:itc} \\
C_{fo} \frac{\partial^2 \Omega_{fo}}{\partial t^2}  &= - \frac{\{B^2_s\} }{\rho \mu} \left(c_f \frac{\partial \Omega_f}{\partial s} \right) \Bigg|_{s=r_i} \nonumber\\
&\quad - \int_{r_i}^{r_f} {\cal F}_m \frac{\partial}{\partial t} \left[ \Omega_f - \Omega_m \right] ds  \, . \label{eq:otc} 
\end{align}
\end{subequations} 
As eq. \eqref{eq:otc} shows, in the absence of coupling at the CMB (${\cal F}_m = 0$), a change in the net angular momentum of the ROTC can only occur as a result of a Lorentz torque at the TC.  The equal and opposite torque applies on the RITC.

The combination of eqs. \eqref{eq:tqm}-\eqref{eq:dtalpha} and eqs. \eqref{eq:itc}-\eqref{eq:otc} captures the coupled oscillation between the mantle, inner core, RITC and ROTC that defines the CMG mode. It involves three torques of restoring nature: the gravitational torque between the inner core and mantle; the Lorentz torque at the TC between the RITC and ROTC, and the EM torque at the ICB between the inner core and the RITC.  The latter (through ${\cal F}_i$) includes also a dissipative part, and EM coupling at the CMB and viscous relaxation of the inner core also contribute to dissipating the mode energy.

Building an analytical prediction of the period and decay rate of the CMG mode as a function of model parameters is not straightforward because  $\{ B^2_s\}$, ${\cal F}_i$ and ${\cal F}_m$ all vary with $s$, and  the profile of $\Omega_f(s)$ is expected to change with the phase of the mode. Our chief objective, instead, is to identify the key factors controlling the oscillating dynamics of the CMG mode.  For this purpose, we can prescribe a simple form of $\Omega_f(s)$. For oscillation periods much longer than the travel time of Alfv\'en waves, the latter should act to redistribute any local change in angular velocity toward a mean $\Omega_f(s)$ profile.  This profile should approach $\Omega_i$ near $s=0$ (because of the expected strong EM coupling at the ICB) and approach $\Omega_{fo}$ in the ROTC. Let us write 
\begin{equation}
\Omega_f(s) = \Omega_i + f(s) (\Omega_{fo}- \Omega_i) \, . \label{eq:omegaf}
\end{equation}
The function $f(s)$ should tend to zero as $s \rightarrow 0$ and approach 1 in the ROTC.  $f(s)$ is time-dependent, but the form of eq. \eqref{eq:omegaf} is sufficiently general to capture the profile of $\Omega_f(s)$ in the CMG mode. 

EM coupling at the CMB and viscous inner core relaxation act to dissipate the energy of the CMG mode. Their effects are important, as we show in our results.  However, to understand the CMG mode dynamics, we want first to focus on the factors that contribute to sustaining it.  For this purpose, let us assume no EM coupling at the CMB (${\cal F}_m \rightarrow 0$) and a rigid inner core ($\tau_i \rightarrow \infty$).  Using $\Omega_f(s)$ from eq. \eqref{eq:omegaf} in eqs. \eqref{eq:tqm}-\eqref{eq:dtalpha} and eqs. \eqref{eq:itc}-\eqref{eq:otc}, we obtain the following coupled system:
\begin{subequations}
\begin{align}
 C_m \frac{\partial^2 \Omega_m}{\partial t^2}   & =  \overline{\Gamma} \left[ \Omega_i - \Omega_m \right]   \, , \label{eq:tqm2} \\
 C_i  \frac{\partial^2 \Omega_i}{\partial t^2} & = - \overline{\Gamma}  \left[ \Omega_i - \Omega_m \right]  +   {\cal K}_i  \frac{\partial}{\partial t} \left[ \Omega_{fo} - \Omega_i \right]    \, ,\label{eq:tqi2} \\ 
C_{fi} \frac{\partial^2 \Omega_{fi}}{\partial t^2} &=  \Gamma_{tc}  \left[ \Omega_{fo} - \Omega_i \right]  - {\cal K}_i  \frac{\partial}{\partial t} \left[ \Omega_{fo} - \Omega_i \right]  \, , \label{eq:itc2} \\
C_{fo} \frac{\partial^2 \Omega_{fo}}{\partial t^2}  &= - \Gamma_{tc}  \left[ \Omega_{fo} - \Omega_i \right]  \, , \label{eq:otc2} 
\end{align}
\end{subequations}
where
\begin{subequations}
\begin{align}
\Gamma_{tc} &= 4\pi r_i^3 \sqrt{r_f^2-r_i^2} \left( \frac{\{B^2_s\} }{\mu} \frac{\partial f}{\partial s}  \right) \Bigg|_{s=r_i} \, , \label{eq:gammatc} \\
{\cal K}_i  &=  \int_o^{r_i} {\cal F}_i f(s) \, ds \, . \label{eq:calKi}
\end{align}
\end{subequations}

This simplified system captures the essential ingredients of the CMG mode. The mode frequency and the relative phase of each of the four regions depend on $\overline{\Gamma}$, $\Gamma_{tc}$ and ${\cal K}_i$.  At leading order, the CMG mode is an exchange of angular momentum between the mantle and the ROTC (see eq. \ref{eq:angmom_approx}) so $\Omega_m$ and $\Omega_{fo}$ should oscillate in phase but in opposite direction.  The leading order dynamics of the CMG mode can thus be illuminated by forming an equation for the differential angular velocity $\Omega_{fo}-\Omega_m$. We build this equation by multiplying eqs. \eqref{eq:otc2} and \eqref{eq:tqm2} by $C_m$ and $C_{fo}$, respectively, and subtracting the two resulting equations.  This gives,
\begin{align}
& C_m C_{fo}  \frac{\partial^2}{\partial t^2}  \left[\Omega_{fo} - \Omega_m\right] = \nonumber\\
& \qquad \qquad - C_{fo} \overline{\Gamma} \left[\Omega_i - \Omega_m\right] - C_m \Gamma_{tc}  \left[\Omega_{fo} - \Omega_i \right] \, . \label{eq:freqapprox}
\end{align}
The two torques on the right-hand side depend on $\Omega_i$.  Whether $\Omega_i$ tends to follow $\Omega_m$ or $\Omega_{fo}$ more closely, or remains somewhere in between $\Omega_m$ and $\Omega_{fo}$, depends on the relative strengths between $\overline{\Gamma}$ and $\Gamma_{tc}$.  A stronger $\overline{\Gamma}$ acts to bring $\Omega_i$ into co-rotation with $\Omega_m$, whereas a stronger $\Gamma_{tc}$ acts instead to limit the differential rotation between $\Omega_{fo}$ and $\Omega_{fi}$, and thus also $\Omega_i$ since it is tied to $\Omega_{fi}$ by EM coupling at the ICB.  This competition between $\overline{\Gamma}$ and $\Gamma_{tc}$ determines the phase of $\Omega_i$ and also the frequency of the CMG mode.  

It is then useful to explore two end-member limits based on the relative coupling strengths between $\overline{\Gamma}$ and $\Gamma_{tc}$.  To give an indication of the expected strength of $\Gamma_{tc}$, taking $\{B_s\} = 3$ mT in eq. \eqref{eq:gammatc}, and approximating $\partial f/\partial s \sim1/r_i$, gives $\Gamma_{tc} \approx 4 \times 10^{20}$ N m.  When $\overline{\Gamma} \ll \Gamma_{tc}$, the strong EM coupling at the TC acts to limit the difference between $\Omega_{fo}$ and $\Omega_{fi}$ (and thus $\Omega_i$).  In this case, the whole of the core rotates as a quasi-rigid body, $\Omega_i$ is in phase with $\Omega_{fo}$ and their amplitudes are similar.  This is the rigid core regime (Figure \ref{fig:cmg_cartoon}b).  Because $\Omega_{fo} \approx \Omega_i$, the frequency of the CMG mode is proportional to $(\overline{\Gamma}/C_m)^{1/2}$ (see eq. \ref{eq:freqapprox}).  When $\overline{\Gamma} \gg \Gamma_{tc}$, $\Omega_i$ is instead locked to $\Omega_m$ by the strong gravitational coupling between the inner core and mantle.  Since $\Omega_{fo}$ is oscillating in the opposite direction as $\Omega_m$, the contrast between $\Omega_{fo}$ and $\Omega_i$ is maximized,  and the frequency of the CMG mode should then be proportional to $(\Gamma_{tc}/C_{fo})^{1/2}$ (see eq. \ref{eq:freqapprox}); this is the locked inner core regime (Figure \ref{fig:cmg_cartoon}c). We further explore each of these two limits in the next two subsections. 

\subsection{The rigid core regime}

In the limit when $\overline{\Gamma} \ll \Gamma_{tc}$, then $\Omega_{fo}\approx \Omega_i\approx \Omega_{fi}$; the whole of the core rotates as a quasi-rigid body.  Let us denote the mean angular velocity of this quasi-rigid core by $\Omega_c$. The CMG mode dynamics is captured by eq. \eqref{eq:tqm2} and an equation formed by summing eqs. \eqref{eq:tqi2}, \eqref{eq:itc2} and \eqref{eq:otc2}, 
\begin{subequations}
\begin{align}
 C_m \frac{\partial^2 \Omega_m}{\partial t^2}   & =  \overline{\Gamma} \left[\Omega_c - \Omega_m\right]   \, , \label{eq:tqm_rc} \\
 C_c  \frac{\partial^2 \Omega_c}{\partial t^2} & = - \overline{\Gamma}  \left[\Omega_c - \Omega_m\right]   \, ,\label{eq:tqi_rc2}  
\end{align}
\end{subequations}
where $C_c = C_{fo}+C_{fi}+C_i$ is the moment of inertia of the whole of the core. Setting $\Omega_m$ and $\Omega_c$ proportional to $\exp(-i \omega_o t)$, the undamped frequency $\omega_o$ of the CMG mode in this limit is 
\begin{equation}
\omega_{o} = \sqrt{\frac{\bar\Gamma C}{C_mC_c}}  \, ,\label{eq:cmg1}
\end{equation}
with $C=C_m +C_c$ the moment of inertia of the whole Earth.  Even though the gravitational torque that sustains the oscillation is purely between the mantle and inner core, the strong EM coupling at both the ICB and the TC entrains the whole of the fluid core into co-rotation with the inner core.   In this limit, the CMG mode is analogous to the MICG mode except that it involves the whole of the core.

The small differential rotation at the ICB implies a weak rate of dissipation by EM coupling at the ICB.  Dissipation in the rigid core regime may then be dominated by EM coupling at the CMB and by the viscous relaxation of the inner core. To re-introduce their influence, we form a coupled system by combining eq. \eqref{eq:tqm} and an equation formed by adding eq. \eqref{eq:tqi} to eq. \eqref{eq:omfapp} integrated from $s=0$ to $r_f$, with the approximation $\Omega_f=\Omega_i=\Omega_c$ and $\partial \Omega_f/\partial s = 0$ (and thus $b_\phi=0$),
\begin{subequations}
\begin{align}
 C_m  \frac{\partial}{\partial t} \Omega_m  &=  \overline{\Gamma} \alpha + {\cal K}_m  \left[ \Omega_c - \Omega_m \right] \, , \label{eq:tqm_rc2} \\
C_c  \frac{\partial}{\partial t} \Omega_c  &= - \overline{\Gamma} \alpha - {\cal K}_m  \left[ \Omega_c - \Omega_m \right]  \, ,\label{eq:tqc_rc2}
\end{align}
\end{subequations}
where 
\begin{equation}
{\cal K}_m  =  \int_o^{r_f} {\cal F}_m \, ds \, .
\end{equation}

Taking $C_c (\partial/\partial t + 1/\tau_i)$ of eq. \eqref{eq:tqm_rc2}, $C_m (\partial/\partial t + 1/\tau_i)$ of eq. \eqref{eq:tqc_rc2}, substituting eq. \eqref{eq:dtalpha}, and subtracting the two resulting equations, we form a damped harmonic oscillator equation for the differential angular velocity $\Delta \Omega = \Omega_c -\Omega_m$,
\begin{equation}
\frac{\partial^2 \Delta \Omega}{\partial t^2}  =- \left( \omega_o^2 + \frac{1}{\tau_i \tau_m} \right) \Delta \Omega - \left( \frac{1}{\tau_i} +  \frac{1}{\tau_m} \right)  \frac{\partial \Delta \Omega}{\partial t} \, , \label{eq:damp_harm}
\end{equation}
where $\tau_m$ is a timescale of attenuation due to EM damping at the CMB given by
\begin{equation}
\tau_m = \frac{C_m C_c}{{\cal K}_m C} \, .\label{eq:taum}
\end{equation}
Setting $\Delta \Omega$ proportional to $\exp(-i \omega t - \lambda t)$, the frequency $\omega$, decay rate $\lambda$ and quality factor ($Q$) of the CMG mode in the rigid core regime are 
\begin{subequations}
\begin{align}
\omega & = \omega_o \left(1 - \left( \frac{\tau_i -\tau_m}{2 \omega_o \tau_i \tau_m} \right)^2 \right)^{1/2}  \, , \label{eq:omega_cmg_rigid} \\
\lambda & = \frac{1}{2} \frac{\tau_i+\tau_m}{\tau_i \tau_m} \, , \label{eq:lambda_cmg_rigid} \\
Q & = \frac{\omega}{2 \lambda} = \left( \left( \frac{ \tau_i \tau_m \omega_o}{\tau_i +\tau_m} \right)^2 - \frac{1}{4}\left( \frac{\tau_i - \tau_m}{\tau_i +\tau_m} \right)^2 \right)^{1/2} \, .\label{eq:Q_cmg_rigid}
\end{align}
\end{subequations}
It may appear paradoxical that the frequency depends solely on $\overline{\Gamma}$ given that the rigid core regime applies when $\overline{\Gamma}\ll \Gamma_{tc}$.  The reason is that since the strong magnetic field locks the whole of the core into a rigid rotation, the magnetic torques at the TC and ICB are much weaker than the gravitational torque between the mantle and inner core.  

To give a numerical example of a typical CMG mode period in this regime, let us take the values of $C_m$ and $C_c$ in Table 1, $\tau_i \rightarrow \infty$, $\tau_m \rightarrow \infty$, and  $\overline{\Gamma} = 10^{19}$ N m;  this gives an undamped CMG period of 182 yr.  To ensure that the CMG mode  is not overdamped ($Q>1$), 
the attenuation timescale $\tau$ formed by
\begin{equation}
\frac{1}{\tau} =  \frac{1}{\tau_m} + \frac{1}{\tau_i}\, ,
\end{equation}
should be longer than $1/\omega = T/2\pi$ where $T$ is the mode period.  To continue the example above, for a CMG mode with a period of $\sim$180 yr, $Q>1$ requires that $\tau_i$ and $\tau_m$ be both longer than 30 yr.

\subsection{The locked inner core regime}

In the limit when $\overline{\Gamma} \gg \Gamma_{tc}$, the inner core is locked to the motion of the mantle, so $\Omega_i=\Omega_m$. 
The CMG mode in this limit is dominated by the oscillation between the RITC and ROTC of eqs. \eqref{eq:itc2} and \eqref{eq:otc2}. The dynamics of the ROTC can be retrieved from eq. \eqref{eq:otc2} and an equation formed by adding eqs. \eqref{eq:itc2} and \eqref{eq:otc2}, using $C_{fi} \ll C_{fo}$ and setting $\Omega_i=\Omega_m$: 
\begin{subequations}
\begin{align}
C_{fo} \frac{\partial^2 \Omega_{fo}}{\partial t^2}  &= - \Gamma_{tc}  \left[ \Omega_{fo} - \Omega_m \right]  \, ,  \label{eq:otc_freq} \\
C_{fo} \frac{\partial^2 \Omega_{fo}}{\partial t^2}  &=  - {\cal K}_i  \frac{\partial}{\partial t} \left[ \Omega_{fo} - \Omega_m \right]  \, . \label{eq:otc_dissisp} 
\end{align}
\end{subequations}
Using $\Omega_m/\Omega_{fo} = - C_{fo}/C_m$  from eq. \eqref{eq:angmom_approx}, and setting $\Omega_{fo}$ and $\Omega_m$ proportional to $\exp(-i \omega t - \lambda t)$ gives the following predictions for $\omega$, $\lambda$ and $Q$: 
\begin{subequations}
\begin{align}
\omega^2 & = \frac{\Gamma_{tc}}{C_{fo}} \left(1 + \frac{C_{fo}}{C_m} \right) \, , \label{eq:omega_cmg_locked} \\
\lambda & = \frac{Re[{\cal K}_i]}{C_{fo}}  \left(1 + \frac{C_{fo}}{C_m} \right)\, , \label{eq:lambda_cmg_locked} \\
Q & = \frac{\omega}{2\lambda} =  \frac{\sqrt{\Gamma_{tc} C_{fo} C_m}}{2 Re[{\cal K}_i] \sqrt{ C_{fo} + C_m}} \, . \label{eq:Q_cmg_locked}
\end{align}
\end{subequations}
These predictions depend on $\Gamma_{tc}$ and ${\cal K}_i$, which in turn depend on $\{B_s\}$ at the TC and on the profile of $\Omega_f(s)$.  For the purpose of computing estimates of $\omega$, $\lambda$ and $Q$, let us assume a simple profile based on eq. \eqref{eq:omegaf} with $f(s)$ given by the following sinusoidal function,
\begin{subequations}
\begin{align}
f(s) & = \frac{1}{2} \left[1 + \sin \left(\frac{\pi}{r_i}(s-r_i) \right) \right] \, , \hspace*{0.5cm} \frac{r_i}{2}  < s < \frac{3 r_i}{2}  \, , \label{eq:fsapprox}\\
f(s) & = 0  \, , \hspace*{0.5cm}  s<\frac{r_i}{2}  \, , \\
f(s) & = 1  \, , \hspace*{0.5cm}  s>\frac{3r_i}{2}\, .
\end{align}
\end{subequations}
For this profile, $\partial f/\partial s$ at the TC is equal to $\frac{1}{2}\pi/r_i$, and taking a representative $\{B_s\}=3$ mT at the TC gives $\Gamma_{tc} = 6.87 \times 10^{20}$ N m.  Using $C_m$ and $C_{fo}$ from Table 1, this gives a period of 21.7 yr. Using this period to compute the magnetic skin depth at the ICB, on which ${\cal F}_i$ depends (see eqs. 5-6 of Paper 1), and integrating the product of ${\cal F}_i$ and $f(s)$ inside the TC (eq. \ref{eq:calKi}) gives $Re[{\cal K}_i ] = 1.99 \times 10^{28}$ kg m$^2$ $s^{-1}$, which gives $Q=1.88$.  These estimates of the period and $Q$ are crude as they entirely depend on our assumed choice of $f(s)$. The true profile of $f(s)$ differs from this simple function, depends on the choice of model parameters and also varies with the phase of the mode.  Nevertheless, we will show that these estimates of $\omega$ and $Q$ are correct in an order of magnitude sense.  Note also that the frequency in the locked inner core regime is independent of $\overline{\Gamma}$; this is expected since  $\overline{\Gamma}$ is sufficiently strong to lock the mantle and inner core into co-rotation. 

EM coupling at the CMB is not included in eqs. \eqref{eq:otc_freq}-\eqref{eq:otc_dissisp} but further contributes to attenuate the CMG mode.  Also note that since the inner core and mantle are locked into co-rotation, $\alpha$ (the angle of longitudinal misalignment, eq. \ref{eq:dtalpha}) remains small and viscous inner core relaxation is then not an important source of dissipation in the locked inner core regime.

\section{Results}

\subsection{The CMG mode in the spectrum of modes}

We now compute the free modes of axial oscillations of the coupled system of eqs. (\ref{eq:tqm}-\ref{eq:dtalpha}) and eqs. (\ref{eq:omfapp}-\ref{eq:bphiapp}).  These are found using the same numerical technique as described in Paper 1.  We assume that each variable has a time-dependency of the form $\exp(-i\omega t)$, so the system can be cast as an eigenvalue problem with eigenvalues $\omega$ corresponding to the complex frequencies of the free modes.  Solutions depend on the chosen profile of $\{ B_s\}$ in the core and on the models of EM coupling at the CMB and ICB. We use the same EM coupling models with the same numerical values as in Paper 1, although we use different choices of mantle conductance.  We also use the same $\{ B_s\}$ profile, given by 

\begin{equation}
\{ B_s \} = \overline{B}_s \sin \left(\frac{\pi s}{r_f} \right) + \left(\frac{s}{r_f} \right)  \langle B_{r,m} \rangle\, ,
\end{equation} 
with $\overline{B}_s=3.3$ mT and $\langle B_{r,m} \rangle=0.319$ mT, with the latter representing the rms strength of the total radial field at the CMB.  Note that for this profile $\{ B_s \} = 0$ at $s=0$, even though in reality the non-axisymmetric background field should contribute to a non-zero part.  This, however, has little influence on our results as cylinders close to the rotation axis are forced into co-rotation with the inner core due to the strong EM coupling at the ICB. A summary of all the numerical parameters used to produce our results is given in Table \ref{tab:params}.

EM coupling at the ICB depends on the magnetic skin depth $\delta=\sqrt{2\eta/|\omega|}$, which depends on the frequency.  The modes that we compute are obtained using the following procedure.  First, we set $\delta$ based on a value of $\omega = 2\pi/6$ yr$^{-1}$ so that the first few harmonics of TO modes are computed approximately correctly for our assumed $\overline{B}_s$.  The CMG mode though has a lower frequency. To improve its accuracy, we use an iterative procedure, updating $\delta$ with the frequency of the CMG mode found at the previous iteration.  Convergence is achieved after three or four iterations. 

\begin{table}[h]
\caption{Parameters used in calculations of the normal modes. \label{tab:para}} \vspace*{0.5cm}
\label{tab:params}
\begin{tabular}{@{}ll} \hline
Parameter & value \\
\hline
radius of the core & $r_f = 3.480 \times 10^6$ m \\
radius of the inner core & $r_i = 0.35 r_f = 1.218 \times 10^6$ m \\
moment of inertia, mantle & $C_m = 7.13 \times 10^{37}$ kg m$^{2}$ \\
moment of inertia, inner core & $C_i = 5.87 \times 10^{34}$ kg m$^{2}$ \\
moment of inertia, RITC & $C_{fi} = 2.06 \times 10^{35}$ kg m$^{2}$ (*)\\
moment of inertia, ROTC & $C_{fo} = 9.22 \times 10^{36}$ kg m$^{2}$ (*)\\
moment of inertia, whole core & $C_c = 9.48 \times 10^{36}$ kg m$^{2}$ (*)\\
moment of inertia, whole Earth & $C = 8.08 \times 10^{37}$ kg m$^{2}$ (*)\\
uniform density, fluid core & $\rho = 1.1 \times 10^4$ kg m$^{-3}$ \\
conductivity, core & $\sigma = 10^6$ S m$^{-1}$ \\
magnetic diffusivity, core & $\eta = \frac{1}{\mu \sigma} = 0.796$ m$^2$ s$^{-1}$ \\
kinematic viscosity, fluid core & $\nu = \frac{\eta}{100} = 0.00796$ m$^2$ s$^{-1}$ \\
$s$-magnetic field amplitude, fluid core & $\overline{B}_s  = 3.3$ mT \\
rms radial magnetic field, CMB & $\langle B_{r,m} \rangle = 0.319$ mT \\
axial dipole, CMB & $B_m^d = 0.391$ mT \\
axial dipole, ICB & $B_i^d = 3.0$ mT or $6.0$ mT \\
\hline 
\end{tabular} \\
(*) Based on a uniform fluid core density of $\rho = 1.1 \times 10^4$ kg m$^{-3}$. 
\end{table}

 For our first set of results, we use a dipole field amplitude at the ICB of $B_i^d=3$ mT, a viscous relaxation time for the inner core of $\tau_i = 1000$ yr, and a mantle conductance of $G_m=10^5$ S. The latter choice gives an EM attenuation time of $\tau_m =17100$ years.  Although both $\tau_i$ and $\tau_m$ are much  higher than expected for Earth, our first objective is to demonstrate that the characteristics of the CMG mode are captured by the simplified system of eqs. \eqref{eq:tqm2}-\eqref{eq:otc2} when the main source of dissipation is from EM coupling at the ICB.  More realistic choices of $\tau_i$ and $\tau_m$ are investigated further below. 

\begin{figure}[ht!]
  \includegraphics[width=8.6cm]{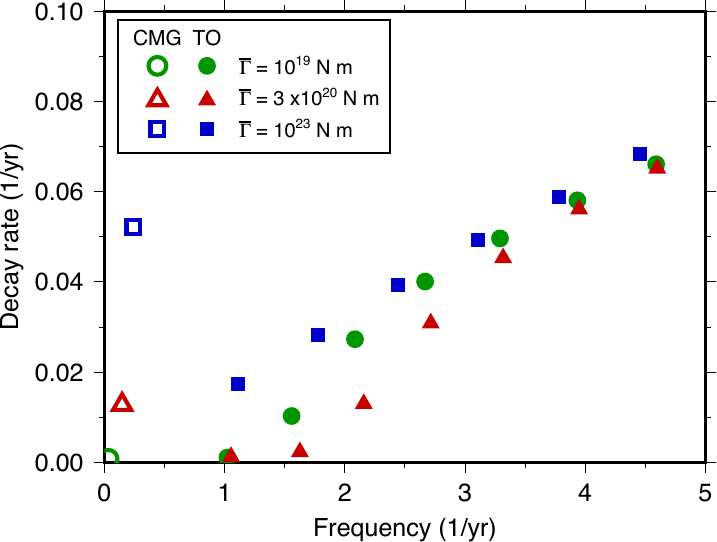} 
  \caption{Frequency and decay rate of a subset of the normal modes computed with $\overline{\Gamma}= 10^{19}$ N m (green circles), $\overline{\Gamma}=3 \times 10^{20}$ N m (red triangles) and $\overline{\Gamma}=10^{23}$ N m (blue squares).  Full symbols denote TO modes, open symbols denote the CMG modes.} 
    \label{fig:freqdecay}
\end{figure}

Figure \ref{fig:freqdecay} shows the frequency and decay rate of a subset of modes computed for three different choices of the gravitational strength factor: $\overline{\Gamma}=10^{19}$ N m (green circles), $\overline{\Gamma}=3 \times 10^{20}$ N m (red triangles) and $\overline{\Gamma}=10^{23}$ N m (blue squares).  In each case, the spectrum of modes includes a set of TO modes with a first harmonic that has a period of approximately 6 years (frequency $\sim 1$ yr$^{-1}$).  We also find a mode that has a lower frequency; this corresponds to the CMG mode.  In increasing order of $\overline{\Gamma}$, the periods are 184.9 yr, 42.6 yr and 26.1 yr and the $Q$ values are 21.7, 5.76 and 2.30.

\begin{figure}[ht!]
  \includegraphics[width=8.6cm]{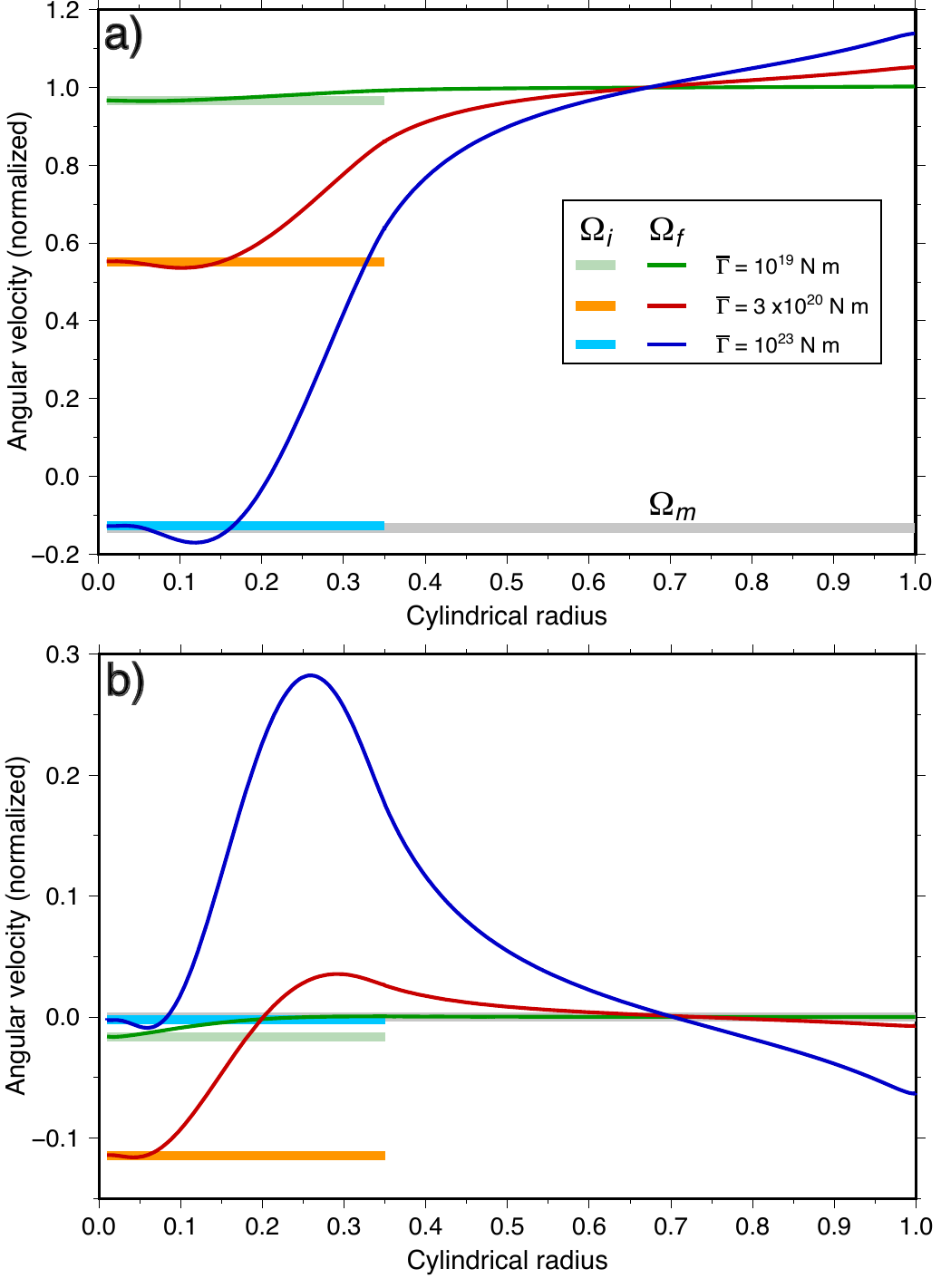} 
  \caption{(a) Real and (b) imaginary part of the angular velocity structure as a function of cylindrical radius ($s/r_f$) for the CMG mode computed with $\overline{\Gamma}=10^{19}$  N m (green line for $\Omega_f$, thick light green line for $\Omega_i$), $\overline{\Gamma}=3 \times 10^{20}$  N m (red lines for $\Omega_f$, thick orange line for $\Omega_i$) and $\overline{\Gamma}=10^{23}$  N m (dark blue line for $\Omega_f$, thick light blue line for $\Omega_i$).  For each choice of $\overline{\Gamma}$, the phase corresponds to when the real part of $\Omega_m$ (thick grey line) is maximum in the retrograde direction (the imaginary part of $\Omega_m$ is equal to zero at that phase).  Amplitudes are normalized such that the real part of the mean angular velocity of the rigid core $\Omega_c = - \Omega_m \cdot (C_m/C_c)$ = 1.} 
    \label{fig:cmgprofiles}
\end{figure}

Figure \ref{fig:cmgprofiles} shows the real and imaginary parts of $\Omega_i$, $\Omega_m$ and the profiles $\Omega_f(s)$ of the CMG modes of Figure \ref{fig:freqdecay}. In each case, the phase is chosen such that the real part of $\Omega_m$ (thick grey line) is maximum in the retrograde direction and the amplitude is normalized such that the real part of the mean angular velocity of the whole core $\Omega_c ( = - \Omega_m \cdot (C_m/C_c) )$ is set equal to 1.  The imaginary part (panel b) is equivalent to the solution obtained a quarter of a cycle later (i.e. with a phase delay of $\pi/2$.) The profile obtained with $\overline{\Gamma} = 10^{19}$ N m illustrates a case when the CMG mode is in the rigid core regime; the inner core follows the motion of the ROTC and the whole core oscillates as a quasi-rigid body.  In contrast, the profile with $\overline{\Gamma} = 10^{23}$ N m is for a case in the locked inner core regime; the motion of the inner core is gravitationally locked to the mantle and oscillates in the opposite direction to the ROTC.  The profile obtained with $\overline{\Gamma} = 3 \times 10^{20}$ N m captures a case in between these two asymptotic regimes, with the  inner core pulled approximately equally by the gravitational torque and by the Lorentz torque at the TC (via EM coupling at the ICB).  For each choice of $\overline{\Gamma}$, the profile of $\Omega_f(s)$ matches $\Omega_i$ near $s=0$ and its mean value in the ROTC is approximately equal to the mean motion of the whole core.  At the phase of the mode when the difference between $\Omega_m$ and $\Omega_{fo}$ is maximum (panel a), the profile of $\Omega_f(s)$ is consistent with the form of eq. \eqref{eq:omegaf}, although this profile is time-dependent.  These solutions also illustrate how the fast propagating Alfv\'en waves in the fluid core act to eliminate large gradients in $\Omega_f(s)$ outside the TC on the longer timescale of the CMG mode; the whole of the ROTC oscillates as a quasi-rigid body.   Note the large differential velocity between $\Omega_f(s)$ and $\Omega_i$ in the cylindrical radius range of $0.1<s/r_f <0.35$; this causes significant EM damping at the ICB (through ${\cal K}_i$).

\subsection{The period and Q of the CMG mode}

\begin{figure}[ht!]
  \includegraphics[width=8.6cm]{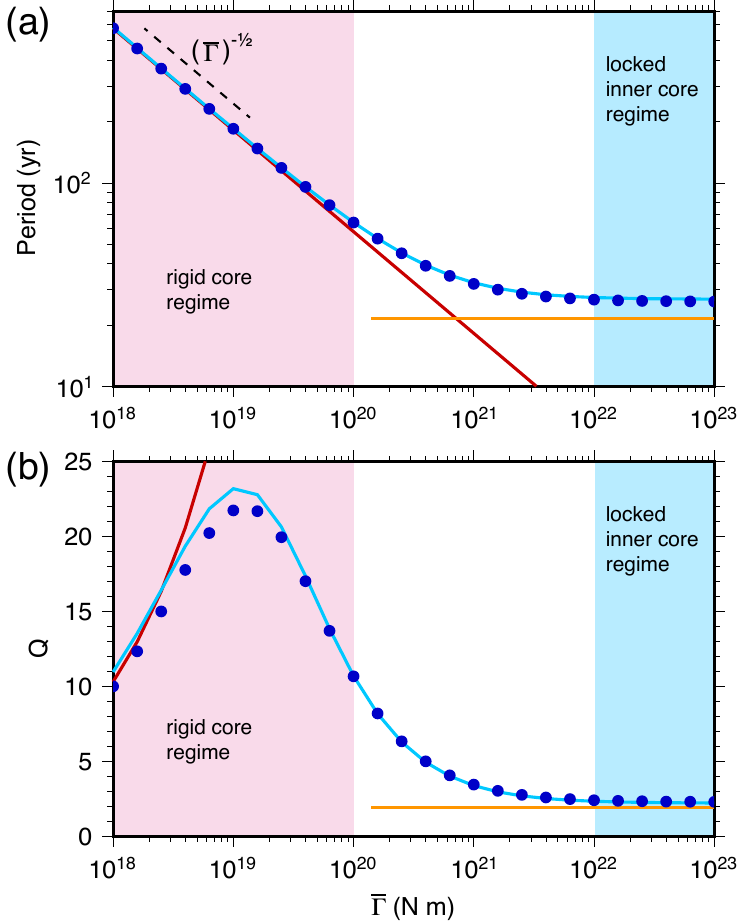} 
  \caption{(a) Period and (b) $Q$ of the CMG mode (blue circles) as a function of $\overline{\Gamma}$. Light blue lines show the predicted periods and $Q$ based on eqs. \eqref{eq:omega_cmg} and \eqref{eq:lambda_cmg}, respectively. Red lines show the predictions in the rigid core regime based on eqs. \eqref{eq:omega_cmg_rigid} and \eqref{eq:Q_cmg_rigid} . Orange lines show the predictions in the locked inner core regime based on eqs. \eqref{eq:omega_cmg_locked} and \eqref{eq:Q_cmg_locked}.   The pink and blue shaded areas highlight the rigid core and locked inner core regimes, respectively. }
    \label{fig:periodQ}
\end{figure}

Figure \ref{fig:periodQ} shows how the period and $Q$ of the CMG mode change with $\overline{\Gamma}$ using the same values of $B_i^d=3$ mT, $\tau_i=1000$ yr, and $G_m=10^5$ S as in the previous subsection. When $\overline{\Gamma} <10^{20}$ N m, the CMG mode is in the rigid core regime, its period (Figure \ref{fig:periodQ}a) is longer than 70 years and is proportional to $(\overline{\Gamma})^{1/2}$.  The period matches very well the prediction from eq. \eqref{eq:omega_cmg_rigid}.  When $\overline{\Gamma} >10^{22}$ N m, the CMG mode is in the locked inner core regime and its period is fixed at 26.1 yr, independent of $\overline{\Gamma}$, in general agreement with the prediction of 24.6 yr based on eq. \eqref{eq:omega_cmg_locked} with $f(s)$ approximated as eq. \eqref{eq:fsapprox}.  The boundaries of  the rigid core and locked inner core regimes are, of course, not precisely defined and the choices adopted on Figure \ref{fig:periodQ} are only meant as a representative guide.

For $\tau_m, \tau_i \rightarrow \infty$, the dynamics of the CMG mode is captured by the reduced system of eqs. \eqref{eq:tqm2}-\eqref{eq:otc2}.  A prediction for its frequency can be built from eq. \eqref{eq:otc2}: 
\begin{equation}
\omega^2  = \frac{\Gamma_{tc}}{C_{fo}} \frac{(\Omega_{fo} - \Omega_i)  }{\Omega_{fo}}  \, .\label{eq:omega_cmg} 
\end{equation}
To first order, $\omega$ depends on $\sqrt{\Gamma_{tc}}$ and on the contrast between $\Omega_{fo}$ and $\Omega_i$; a larger contrast leads to a faster frequency. The amplitude of $\Gamma_{tc}$ varies over the CMG cycle because the profile of $\Omega_f(s)$ is time-dependent (see Fig. \ref{fig:cmgprofiles}). 
We show on Figure \ref{fig:periodQ}a the predicted period computed from eq. \eqref{eq:omega_cmg}, with $\Gamma_{tc}$ (eq. \ref{eq:gammatc}) evaluated from the actual $\Omega_f(s)$ profile of the CMG mode solution at the phase when the difference between $\Omega_m$ and $\Omega_{fo}$ is maximum.  The very good match with the computed modes, for all $\overline{\Gamma}$, confirms that the period of the CMG mode is always consistent with the Lorentz torque at the TC between the RITC and ROTC.  When the RITC is almost co-rotating with the ROTC (rigid core regime), the stretching of the $\{B_s\}$-field is minimal, the Lorentz torque at the TC is weak, and the CMG period is long.  As $\overline{\Gamma}$ is increased, the gravitational pull on the inner core by the mantle is stronger, increasing the differential rotation and the Lorentz torque at the TC, resulting in a shorter CMG mode period. When the differential rotation at the TC is maximized, in the locked inner core regime, the Lorentz torque and CMG period no longer change with increasing $\overline{\Gamma}$.

Figure \ref{fig:periodQ}b shows how $Q$ changes as a function of $\overline{\Gamma}$.  $Q$ increases with $\overline{\Gamma}$ in the rigid core regime,  reaching a maximum of approximately 22 at $\overline{\Gamma}\approx 10^{19}$ N m, then decreases with $\overline{\Gamma}$ before settling to a value of 2.3 in the locked inner core regime. We also show on Figure \ref{fig:periodQ}b the predictions of $Q$ from eq. \eqref{eq:Q_cmg_rigid} in the rigid core regime and from eq. \eqref{eq:Q_cmg_locked} in the locked inner core regime. Both provide reasonable fits to the computed values of $Q$.  
 We can build a prediction of the decay rate $\lambda$ valid for all $\overline{\Gamma}$ (when $\tau_m, \tau_i \rightarrow \infty$) by adding eqs. \eqref{eq:itc2} and \eqref{eq:otc2} and using $C_{fi} \ll C_{fo}$,
  \begin{equation}
\lambda  = \frac{Re[{\cal K}_i]}{C_{fo}} \frac{(\Omega_{fo} - \Omega_i)  }{\Omega_{fo}} \, , \label{eq:lambda_cmg}
\end{equation}
with ${\cal K}_i$ computed from the actual profile of $\Omega_f(s)$ at the phase of the mode when the difference between $\Omega_m$ and $\Omega_{fo}$ is maximum. The prediction of $Q=\omega/2\lambda$ based on this decay rate and $\omega$ from eq. \eqref{eq:omega_cmg} is shown in Figure \ref{fig:periodQ}b;  it provides a good match to our computed values.  This confirms that -- when $\tau_i$ and $\tau_m$ are long -- EM damping at the ICB regulates the dissipation of the CMG mode.  In the rigid core regime, EM damping at the ICB is minimal since the inner core co-rotates with the RITC, and the prediction from eq. \eqref{eq:lambda_cmg} is not as accurate.  At the lowest values of $\overline{\Gamma}$, attenuation is instead primarily from the combination of $\tau_i$ and $\tau_m$, even for the large values that we have used in Figure  \ref{fig:periodQ}b, and $Q$ matches better the prediction of eq. \eqref{eq:Q_cmg_rigid}.  Although $\tau_i$ and $\tau_m$ are fixed, the increase in period with decreasing $\overline{\Gamma}$ implies a lower $Q$ as $\overline{\Gamma}$ gets smaller.  
 
\begin{figure}[ht!]
  \includegraphics[width=8.6cm]{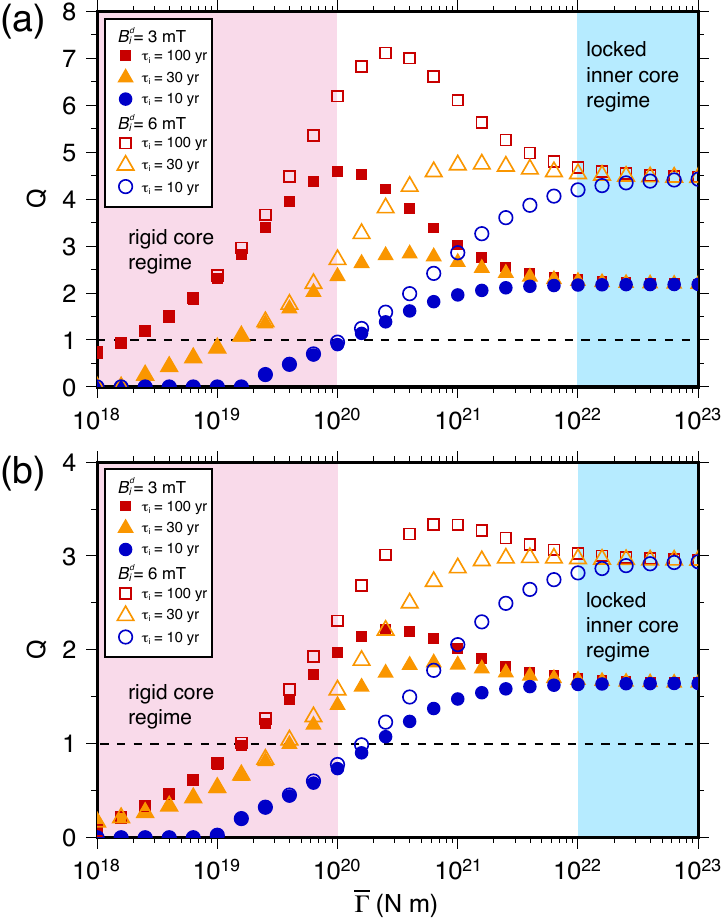} 
  \caption{$Q$ of the CMG mode as a function of $\overline{\Gamma}$ for a mantle conductance of (a) $G_m=7\times 10^6$ S and (b) $G_m = 5 \times 10^7$ S, different choices of viscous relaxation time ($\tau_i=100$ yr, red squares; $\tau_i=30$ yr, orange triangles; $\tau_i=10$ yr, blue circles), and different dipole field amplitude at the ICB (full symbols, $B_i^d = 3$ mT; empty symbols, $B_i^d = 6$ mT). Values of $Q=0$ indicate cases where the CMG mode was not found. The pink and blue shaded areas highlight the rigid core and locked inner core regimes, respectively.  The dashed horizontal line marks $Q=1$. }
    \label{fig:Q}
\end{figure}

While the choices $\tau_i=1000$ years and $G_m=10^5$ S are useful to demonstrate the main properties of the CMG mode, they are not geophysically realistic.  The viscous relaxation time of the inner core is likely much smaller than a 1000 years given that the temperature at the ICB is at the melting point.  Likewise, a more realistic mantle conductance is of the order of $G_m=7 \times 10^6$ S \citep[][]{kuvshinov21}, but it can be higher if there is a thin layer of iron-enriched material at the base of the mantle \citep[][]{ohta12,ohta14,ho24}.  Figure \ref{fig:Q} shows how $Q$ of the CMG mode varies with $\overline{\Gamma}$ for different choices of mantle conductance ($G_m=7 \times 10^6$ S and $5 \times 10^7$ S, corresponding to $\tau_m=244$ yr and   $34$ yr, respectively), inner core viscous relaxation times ($\tau_i=100$ yr, $30$ yr and $10$ yr) and dipole field amplitude at the ICB ($B_i^d=3$ mT and 6 mT). The high value of $B_i^d=6$ mT for the dipole field amplitude at the ICB is not necessarily realistic, but it is used here to illustrate how a stronger radial field results in a tighter EM coupling at the ICB, and thus a weaker EM dissipation and a larger $Q$.  

Lower values of $\tau_i$ and $\tau_m$ increase the decay rate and lead to a lower $Q$. Depending on the specific choices of $G_m$, $\tau_i$ and $B_i^d$, $Q$ values on Figure \ref{fig:Q} range between 0 and 7.  Smaller values of $\tau_i$ and larger values of $G_m$ (smaller $\tau_m$) lead to a smaller $Q$, as expected.  A larger $B_i^d$ leads to larger $Q$.  In the rigid core regime ($\overline{\Gamma}<10^{20}$ N m), $Q$ is independent of $B_i^d$ as the inner core co-rotates with the fluid core and EM dissipation at the ICB is minimized. For $\overline{\Gamma}<10^{19}$ N m, $Q$ is small (typically below 1) because the period of the CMG mode is long compared to $\tau_i$ and $\tau_m$.  In the locked inner core regime ($\overline{\Gamma}>10^{22}$ N m), $Q$ is independent of $\tau_i$ (as expected), but depends on EM dissipation at both the CMB (a larger $G_m$ leads to a smaller $Q$) and the ICB (a weaker $B_i^d=3$ mT leads to a smaller $Q$).  So long as $Q$ is of order 1 or larger, the period of the CMG mode is not significantly altered from those shown in Figure \ref{fig:periodQ}a. 
 
\section{Discussion and conclusions}

We have presented the attributes of an axial mode of oscillation of the coupled core-mantle system sustained by the gravitational torque between the mantle and inner core which we have termed the CMG mode.  In contrast to the MICG mode, which involves an oscillation between the mantle and inner core (whether entraining or not the fluid region inside the TC), the CMG mode involves the whole of the core.  It consists in an exchange of angular momentum primarily between the mantle and the fluid region outside the TC, with the latter oscillating quasi-rigidly as a result of strong internal magnetic coupling.  The core-mantle exchange of angular momentum occurs through a set of three internal torques.  First, the fluid outside the TC transmits angular momentum to the fluid region inside the TC by a magnetic torque.  Second, the fluid inside the TC transmits angular momentum to the inner core by EM coupling at the ICB.  Finally, the inner core transmits angular momentum to the mantle by a gravitational torque.  This form of gravitational oscillation, involving the whole of the core instead of only the inner core, occurs when the magnetic field within the core is sufficiently strong that the propagation time of Alfv\'en waves is shorter than the mode period, which is the case for Earth.

The behaviour of the inner core differs in the CMG and MICG modes. In the MICG mode the angular velocity of the inner core is opposite that of the mantle, and larger by a factor $C_m/(C_i+C_{fi})$. In the CMG mode, the motion of the inner core depends on the relative strengths between $\overline{\Gamma}$ and $\Gamma_{tc}$.  When $\overline{\Gamma}\ll \Gamma_{tc}$, the strong magnetic coupling at the TC entrains the fluid inside the TC and the inner core into co-rotation with the fluid outside the TC.  The whole of the core oscillates as a quasi-rigid body (the rigid core regime) so the angular velocity of the inner core is opposite that of the mantle, just as in the MICG mode, however its amplitude is smaller (approximately equal to $C_m/C_c$ that of the mantle).   When $\overline{\Gamma}\gg \Gamma_{tc}$, the strong gravitational coupling instead locks the inner core into a co-rotation with the mantle (the locked inner core regime).  The behaviour of the inner core is in between these two asymptotic cases when $\overline{\Gamma}$ is of the same order of magnitude as $\Gamma_{tc}$. 

For Earth, assuming a magnetic field $\{B_s\}$ inside the core of approximately 3 mT \citep{gillet10} as we have used, and for the range of $\overline{\Gamma}=[0.3 -2 ] \times 10^{20}$ N m estimated by \citet{davies14}, the CMG mode period should be in the range of 40-100 years (Figure \ref{fig:periodQ}a), either in, or at the edge of, the rigid core regime.  The dissipation from EM damping at the ICB, by itself, should lead to a $Q$ in the range of 5 to 20 (Figure \ref{fig:periodQ}b).  However, EM damping at the CMB and viscous relaxation of the inner core likely reduce $Q$ to lower values.  For a mantle conductance of $7 \times 10^{6}$ S, based on the mantle conductivity model of \citet{kuvshinov21}, and a rms radial magnetic field strength at the CMB of 0.4 mT, the EM attenuation time at the CMB should be of the order of $\tau_m \approx 250$ yr. Yet, EM damping at the CMB may be more significant. Several studies have shown that (Mg,Fe)O becomes highly conducting ($10^4-10^5$ S/m) in the lowermost mantle \citep[e.g.][]{ohta12,ohta14,ho24} and a thin (1 to 2 km) layer of such iron-enriched material at the base of the mantle is supported by normal mode seismology \citep[][]{russell23}.  Indeed, the presence of such a layer provides the conductance required to explain (by EM damping) the observed dissipation in Earth's nutations \citep[e.g.][]{buffett92,buffett02,koot10,koot13}, and the observed attenuation of Alfv\'en waves \citep[][]{gillet10,gillet17}. With a higher conductance of $5 \times 10^{7}$ S, $\tau_m \approx 34$ yr, and the CMG mode should then be rapidly attenuated.  In addition, viscous relaxation of the inner core also acts to attenuate the CMG mode.  As we have shown (Figure \ref{fig:Q}b), with a mantle conductance of $5 \times 10^{7}$ S, and for $\overline{\Gamma} = [0.3 -2] \times 10^{20}$ N m, the viscous relaxation time of the inner core $\tau_i$ has to be longer than 10 yr for $Q>1$.  Using the mapping provided by \citet{buffett97}, this corresponds to an inner core viscosity of $5 \times10^{17}$ Pa s.  The viscosity of the inner core is not well known, but inferences from seismic analyses suggest that the inner core topography deforms on a relatively short time, which implies a viscosity in this range \citep{vidale25}. To summarize, the attenuation from both EM coupling at the CMB and viscous relaxation of the inner core implies that if $Q$ is above 1, it is likely not significantly larger than 1. 

Given its low expected $Q$, it is then unlikely that a freely oscillating CMG mode is responsible for any specific frequency in the spectrum of the observed decadal fluctuations in the length of day (LOD) \citep[e.g.][]{roberts07}.  These decadal LOD changes are caused by forced zonal accelerations in the fluid core driven by convection \citep[e.g.][]{more18,schwaiger24,aubert26}. The CMG mode may provide a resonant amplification to the zonal accelerations in the core that have a timescale in the multi-decadal range, but this amplification is likely modest at best given the expected low $Q$ of the CMG mode. 

Recent numerical models of the geodynamo that approach Earth's core conditions and include gravitational coupling appear to show evidence of a LOD amplification by the CMG mode \citep[][]{aubert26}.  For the parameters used in the geodynamo model ($\overline{\Gamma} = 5 \times 10^{19}$ N m, $\tau_i = 20$ yr, and three choices of mantle conductance: $2.2 \times 10^7$ S, $2.2 \times 10^8$ S and $2.2 \times 10^9$ S), the CMG mode should be in the rigid core regime with a period of approximately 80 yr.  $Q$ should be above 1 only for the lowest mantle conductance case ($2.2 \times 10^7$ S), and indeed only this case features an amplification of the LOD changes at multidecadal timescales. Whether this amplification indeed results from the CMG mode needs to be verified, but it is in line with our predictions.
  
One limitation of our study is the assumption of axially invariant flow in the fluid core.  Oscillations involving axial flow gradients are possible in the form of Magnetic-Coriolis (MC) modes \citep[][]{dumberry25b}.  Even though the low harmonics of axisymmetric MC modes have longer periods, of the order of a thousand years, the flow structure in the CMG mode may be modified by these MC modes in such a way that some departure from axial rigidity may occur at periods of 50 to 100 years. Indeed, EM coupling at the ICB may trigger such axisymmetric MC waves \citep[][]{dumberry99}, which may then alter the effective strength of this coupling, and also affect the attenuation of the CMG mode. It is thus imperative that the attributes of the CMG mode should be confirmed with a more general model of zonal core flows.  In addition, if the top of the core is stratified, with a buoyancy frequency approximately equal to Earth's rotation rate, free oscillations in the form of Magnetic-Archimedean-Coriolis (MAC) modes may have periods in the same multi-decadal range as the CMG mode \citep{buffett14,buffett16}.  It is then possible that MAC modes interact with the CMG mode, although a dedicated study on this topic is required to understand  the precise form of this interaction.

Finally, even though our study has been anchored to Earth, the ideas of the CMG mode apply in a general sense to any planetary body that has an inner core at its centre and a strong magnetic field permeating its core.

\section*{Appendix A}
\setcounter{table}{0}
\def\thetable{A\arabic{table}}
\setcounter{equation}{0}
\def\theequation{A\arabic{equation}}
\setcounter{figure}{0}
\def\thefigure{A\arabic{figure}}

It is beyond the scope of our study to perform a thorough comparison of the dynamics captured by our 1D model (eqs. \eqref{eq:omfapp} and \eqref{eq:bphiapp}) versus that in a 2D or 3D model.  Nevertheless, we show in this appendix how the TO modes of our 1D model compare with the modes computed in the 2D models of two recent studies.  

We first compare our TO modes with those from the study of \citet[][]{luo22a}.  Their modes are computed in a full sphere (no inner core), assuming an inviscid fluid core ($\nu=0$), and no EM coupling at the CMB. Table \ref{tab:app} (row labelled LJ2D) gives the frequencies and decay rates of the first six TO modes, in units of Alfv\'en time, for their S2 background magnetic field, and for $Le = 10^{-3}$ and $Lu=2 \times 10^3$.  Here, $Le$ and $Lu$ are the Lehnert and Lundquist numbers defined respectively by 

\begin{equation}
Le = \frac{{\cal B}}{\Omega_o r_f \sqrt{\mu \rho}} \, , \quad Lu =  \frac{{\cal B} r_f}{\eta \sqrt{\mu \rho}} \, ,
\end{equation}
where ${\cal B}$ is the characteristic strength of the magnetic field, $\Omega_o$ is Earth's rotation rate, and other parameters are defined in Table 1.  Table \ref{tab:app} also gives the frequencies and decay rates that they compute  based on a 1D model of their system (row labelled LJ1D). This model consists of the coupled equations (in non-dimensional form)

\begin{subequations}
\begin{align}
 \frac{\partial \Omega_f}{\partial t}  & = \frac{1}{s^3 H} \frac{\partial }{\partial s} \left(s^2 H \{B_s b_\phi \} \right)   \, , \label{eq:omfluo22} \\
 \frac{\partial }{\partial t}  \{B_s b_\phi \} & = s \{B_s^2\} \frac{\partial \Omega_f}{\partial s} \nonumber\\ 
 & \hspace*{-0.3cm} + \frac{1}{Lu} \left[  s \frac{\partial^2}{\partial s^2} \left( \frac{ \{B_s b_\phi \} }{s} \right)+ \frac{\partial}{\partial s}  \left(\frac{ \{B_s b_\phi \} }{s}\right) - \frac{\{B_s b_\phi \} }{s^2} \right]  \, , \label{eq:bphiluo22} 
\end{align}
\end{subequations}
where $H=\sqrt{1-s^2}$.  The frequencies of the TO modes of this 1D model are very close to those of the 2D model, however the decay rates are underestimated because the 1D model cannot capture the radial gradients of $b_\phi$ that develop near the CMB.

We first modify our 1D model to one equivalent to that of \citet{luo22a} except that it also includes viscous diffusion.  That is, we use eq. \eqref{eq:omf} with ${\cal F}_m = {\cal F}_i = 0$, in the absence of an inner core, combined with eq. \eqref{eq:bphiluo22} (in the dimensional form of our model). Table \ref{tab:app} (row labelled  LJ1D+diff) gives the frequencies and decay rates for this model when using $\nu = \eta/100$, as we have used for all our results.  The mode frequencies are identical to those of the 1D model of \citet{luo22a} and the decay rates are only slightly enhanced.  This shows that the addition of viscous diffusion, when $\nu = \eta/100$, only weakly affects the results.  We have verified that when $\nu$ is further decreased the results approach those of the LJ1D model.

\begin{table*}[h]
\begin{center}
\caption{The frequency (imaginary part) and decay rate (negative real part) of the first six TO modes computed by \citet{luo22a} in the full sphere for their S2 background magnetic field, based on their 2D model (LJ2D), their 1D model (LJ1D), and their 1D model with the addition of viscous diffusion (LJ1D+diff).  The last row gives the TO modes retrieved with the 1D model of eqs. \eqref{eq:omfapp} and \eqref{eq:bphiapp}, with ${\cal F}_m = {\cal F}_i = 0$, in the absence of an inner core, and using the same background magnetic field. All results are based on $Le=10^{-3}$ and $Lu=2 \times 10^3$ and reported in units of Alfv\'en time.} \vspace*{0.cm}
\label{tab:app}
\begin{tabular}{@{}lcccccc} \hline
model & mode 1 & mode 2 & mode 3 & mode 4 & mode 5 & mode 6 \\
\hline
LJ2D & -0.0528 + 1.03i & -0.112 + 1.91i & -0.168 + 2.62i & -0.176 + 3.10i & -0.322 + 3.44i & -0.469 + 3.88i \\
LJ1D &-0.0232 + 1.04i &-0.078 + 1.92i &-0.124 + 2.63i &-0.145 + 3.10i &-0.281 + 3.46i &-0.419 + 3.90i \\
LJ1D+diff &-0.0238 + 1.04i &-0.080 + 1.92i &-0.126 + 2.63i &-0.147 + 3.11i &-0.284 + 3.46i &-0.422 + 3.90i \\
This study &-0.0308 + 1.04i &-0.082 + 1.92i &-0.129 + 2.63i &-0.149 + 3.11i &-0.285 + 3.46i &-0.424 + 3.90i \\
\hline 
\end{tabular} \\
\end{center}
\end{table*}

\begin{figure*}[ht!]
  \includegraphics[width=0.9\textwidth]{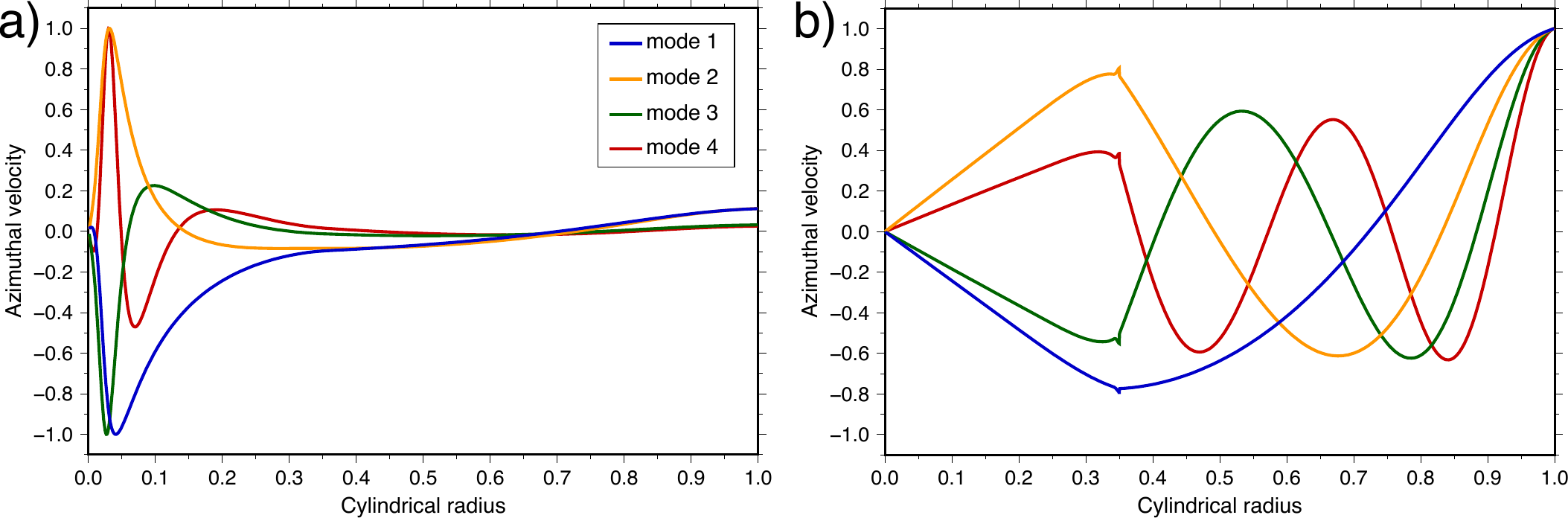} 
    \caption{(a) The structure of the zonal geostrophic flow ($v_\phi$) as a function of cylinder radius ($s/r_f$) for the first four TO modes from our 1D model, when an inner core is present, for no EM coupling at the ICB and CMB, and for a $\{B_s\}$-field based on the SH2 background field of \citet{yuan25}. (b) Modified structure of the modes when EM coupling at the ICB (based on a uniform radial field of $\langle B_{r,i} \rangle =$ 1 mT) is included.  In all cases, the phase of the mode is chosen such that $v_\phi$ at $r_f$ is at its maximum positive value, and the amplitude is scaled such that the largest $|v_\phi(s)|$ equals 1.} 
    \label{fig:app}
\end{figure*}

The last row of Table \ref{tab:app} gives the frequencies and decay rates of the TO modes from our 1D model: that is, the combination of eqs. \eqref{eq:omfapp} and \eqref{eq:bphiapp}, with ${\cal F}_m = {\cal F}_i = 0$, in the absence of an inner core, and using the same (S2) background field and the same values of $Le$ and $Lu$ as \citet{luo22a}. The mode frequencies are identical to those of the LJ1D model. This shows that, in the absence of diffusion (no attenuation), our 1D model is equivalent to that of eqs. \eqref{eq:omfluo22} and \eqref{eq:bphiluo22} used in a number of studies \citep[e.g.][]{jault05,canet09,cox14,maffei16}. The different form of the diffusion term in eq. \eqref{eq:bphiapp} compared to eq. \eqref{eq:bphiluo22} results in a decay rate of mode 1 which is higher than that of model LJ1D, and decay rates for higher modes which are only slightly enhanced compared to LJ1D and LJ1d+diff. Just like for model LJ1D, our 1D model underestimates the magnetic diffusion compared to the 2D model (LJ2D). Overall, our 1D model captures well the frequencies of the TO modes and underestimates their decay rates, in a manner similar to the 1D model of \citet{luo22a}.

We also compare the TO modes of our 1D model with those of the 2D model of \citet{yuan25}, which includes an inner core of radius equal to 0.35$r_f$.  The 2D model of \citet{yuan25} features no coupling at the ICB and CMB, and does not include gravitational coupling.  Fig. \ref{fig:app}a shows the structure of the zonal geostrophic flow $v_\phi$ of the first four TO modes computed from our 1D model when using the background field SH2 of \citet{yuan25} to compute the profile of $\{B_s\}$ and for the same values of $Le=10^{-3}$ and $Lu = 2 \times 10^3$ that they use.  As found by \citet{yuan25}, when an inner core is present, and when the TO modes are decoupled from the mantle and inner core, the largest fluctuations of $v_\phi$ occur inside the tangent cylinder, where $\{ B_s \}$ is weaker and where cylinders have a small moment of inertia.  The TO modes that we compute differ from those of the 2D model of \citet{yuan25} (see their Figure 10): the $v_\phi$ gradients of all modes in our 1D model tend to be more focused toward the rotation axis, and we also do not find a fundamental mode (mode 1) which remains large-scale.  This indicates that important dynamical ingredients inside the TC that act to resist large zonal motion are missing from the 1D model.

However, the confinement of the modes inside the TC is not geophysically realistic for two main reasons.  First, contrary to the simple axisymmetric background field models used by \citet{yuan25}, $\{B_s\}$ should not vanish as $s\rightarrow0$ because the non-axisymmetric parts of the field should contribute to maintaining a non-zero $\{B_s\}$. This should prevent large zonal flow gradients close to the rotation axis (see for example the modes in Figure 2 of \citet{gerick21}, although computed in a full sphere). Second, EM coupling at the ICB should also prevent large zonal flow gradients from developing inside the TC.  Fig. \ref{fig:app}b shows how the structure of the first four TO modes of our 1D model changes when adding EM coupling at the ICB of the form 
\begin{equation}
{\cal F}_i(s) = 4 \pi s^3 G_i  \frac{r_i}{z_i}  \langle B_{r,i} \rangle^2    \, ,\label{eq:fmi}
\end{equation}
where $G_i$ is a conductance factor (Eq. 6 of Paper 1), and with a relatively weak uniform radial field strength of $ \langle B_{r,i} \rangle=1$ mT (much weaker than the mean strength of approximately 30 mT implied by $Le = 10^{-3}$).  The zonal flow inside the TC is restricted to broadly follow the motion of the inner core, preventing a concentration of structures inside the TC, and the TO modes are large-scale, akin to those in the full sphere.  (The small kink in the curves at $s/r_f =0.35$ is caused by the divergence of the EM coupling model as $s \rightarrow r_i$, mitigated by a geometrical adjustment; see Eq. 10 of Paper 1.)  EM coupling at the ICB provides then the necessary resistance to large zonal flow gradients inside the TC, removing the most important discrepancy between the 1D and 2D models.

\begin{acknowledgements}
The suggestions and comments by two reviewers have greatly improved the quality of this paper. This work was financially supported by Discovery Grant RGPIN-2025-05158 from NSERC/CRSNG (Canada).  Figures were produced using the GMT software \citep{gmt}.
\end{acknowledgements}

\section*{Data availability}

Codes to reproduce all numerical experiments are freely accessible at the following data repository: \citet{dumberry26data}.

\section*{Competing interests}  
      
The author of this paper has no competing interests.

\printbibliography

\end{document}